\documentclass[11pt]{article}
\pdfoutput=1
\usepackage{color}
\usepackage{biblatex}
\usepackage{hyperref}

\hypersetup{pdfauthor={Shea Ketsdever and Michael J. Fischer},pdftitle={Incentives
    Don't Solve Blockchain's Problems}}

\title{Incentives Don't Solve Blockchain's Problems}

\author{Shea Ketsdever\\[1ex]
  \small  Department of Computer Science,\\
  \small  Yale University, New Haven, USA\\
  \small\texttt{shea.ketsdever@yale.edu}
  \and Michael J. Fischer\\[1ex]
  \small  Department of Computer Science,\\
  \small  Yale University, New Haven, USA\\
  \small\texttt{michael.fischer@yale.edu}
}

\begin{document}

\maketitle
\begin{abstract}

\noindent A blockchain faces two fundamental challenges. It must motivate users to maintain the system while preventing a minority of these users from colluding and gaining disproportionate control. Many popular public blockchains use monetary incentives to encourage users to behave appropriately. But these same incentive schemes create more problems than they solve. Mining rewards cause centralization in \textit{proof of work} chains such as Bitcoin. Validator rewards and punishments invite attacks in \textit{proof of stake} chains. This paper argues why these incentive schemes are detrimental to blockchain. It considers a range of other systems---some of which incorporate monetary incentives, some of which do not---to confirm that monetary incentives are neither necessary nor sufficient for good user behavior.

\end{abstract}

\section{Introduction}

\noindent A blockchain implements a decentralized and distributed ledger. Users update this ledger by posting \textit{transactions} to a network of their peers, who collect the transactions into batches known as \textit{blocks}. These blocks are sequentially chained together to create a \textit{blockchain}, the record of all accepted transactions \cite{Nakamoto09}.

There may be many copies of the blockchain at different places in the network. The system must propagate changes to these copies across the network to keep all of the blockchains synchronized. The system must also resolve conflicts, such as \textit{forking}, where multiple users add different new blocks to their copies of the chain simultaneously. In such cases, the system must use some mechanism to resolve the conflicts and establish consensus on what the correct ledger is. Most popular blockchains, including Bitcoin and Ethereum, use a longest-chain heuristic to choose between competing copies of the ledger. Users replace their current copy if and only if they discover a longer valid blockchain.

A primary obstacle for establishing fair consensus in public blockchains is the Sybil attack, where users magnify their influence by creating multiple aliases in the system \cite{Douceur02}. These attacks are possible because the accounts are anonymous, making it impossible to tell if the same person is behind many seemingly unique addresses. By coordinating across their multiple accounts, Sybil attackers can control, delay, or even prevent consensus.

\section{Proof of Work}

\noindent Two methods have been introduced to prevent Sybil attacks on blockchains. The first, known as \textit{proof of work}, was proposed in 2008 by person(s) under the pseudonym Shatoshi Nakamoto \cite{Nakamoto09}. Proof of work uses a \textit{pricing function}---a function which is difficult to compute but easy to check---to slow the rate at which the ledger is updated \cite{Dwork93}. Users must compute this function before they can add the next block to the chain. This prevents Sybil attacks because it is equally computationally intensive to compute the function from one account as it is from multiple accounts.

The pricing function introduces a significant cost to using the system. Adding a block to Bitcoin requires specialized and expensive equipment that consumes enormous amounts of electrical energy. Bitcoin mining consumed 2.55 gigawatts of power in 2018, nearly as much as the country of Ireland that same year \cite{DeVries18}.

To offset the anticipated cost, Nakamoto added \textit{mining rewards} to Bitcoin. These rewards are cash incentives (in the Bitcoin currency) paid to the user who computes the pricing function first.

However, mining rewards have had unfortunate side effects. To ease the burden of the high-cost computation, users have begun to collaborate in \textit{mining pools}, where they pool their computational resources to solve the puzzles together. This benefits the users because it ensures a more steady stream of income. But it centralizes control of the Bitcoin ecosystem because individual users no longer act independently.

As of June 2018, over 80 \% of Bitcoin mining was performed by six mining pools, five of which are based in China \cite{Kaiser18}. This geographic and computational centralization undermines the decentralized premise of blockchain technology. It also increases the likelihood of a \textit{51\% attack}, where a single user or coalition of users controls the majority of computational power and therefore the fate of the chain \cite{BitcoinWiki19}.

\section{Proof of Stake}

\noindent Recognizing the energy inefficiencies in proof of work, a newer form of Sybil resistance, known as \textit{proof of stake}, replaces the pricing function with a weighted lottery system \cite{King12}. The lottery selects a user to add the next block with probability proportional to their \textit{stake} in the system. Stake is typically computed based on wealth, though implementations vary. Users who have the most invested in the system therefore have the most control over the chain.

Proof of stake has many advantages over proof of work. Users need not buy expensive mining equipment to participate. The environmental impacts are also low. Major cryptocurrencies like Ethereum \cite{Wood14} are considering converting to stake-based chains \cite{Buterin17}. But proof of stake is not without its problems. It relies on an incentive scheme with its own troubling set of issues.

In proof of work, mining rewards offset the computational cost of using the system. In proof of stake, there is no pricing function to compute. Therefore, there is no computational burden needing to be offset with a reward. Yet proof of stake employs a system of \textit{validator rewards} where dividends are paid to the user who adds the next block.

In other words, proof of stake inherits proof of work's incentive scheme, despite the fact that it does not inherit the structural issues which necessitated that scheme. These incentives are no longer justified by a pragmatic balance of costs and benefits. Rather, they appear to be motivated by a qualitative assumption about human behavior---that people are primarily and reliably motivated by greed.

But catering to greed neglects and undermines other useful behaviors. This produces new issues for blockchain.

\subsection{Nothing at Stake Problem}

\noindent Users want to be confident in the \textit{finality} of their transactions. When a transaction is added to the ledger, it should not be arbitrarily changed or reversed. However, transaction finality is not guaranteed during a fork. The system will only reach consensus on one copy of the chain, abandoning any transactions made on other copies. Users therefore cannot be confident that transactions will be committed until the system reaches consensus. They have an interest in minimizing this uncertainty and reaching consensus quickly.

Validator rewards, however, create a financial incentive to delaying consensus. Users can increase their likelihood of receiving a reward by building on each copy of the blockchain during a fork. This ensures that they have a chance to win the lottery and collect dividends regardless of which copy is chosen. This tactic is economically attractive in proof of stake because there is little cost involved in adding a new block. But it stalls consensus.

When all users build on all competing forks, the system cannot declare a victor. Each fork will continue to develop, perpetuating uncertainty over the true chain. This is called the \textit{nothing at stake problem} \cite{EthereumGithub18}. The financial upside to building on multiple forks makes it lucrative for honest users to delay consensus and undermine transaction finality. Their short-term interest in collecting rewards conflicts with their long-term interest in maintaining the system.

\subsection{Proposed Solutions to Nothing at Stake}

\noindent Some strategies attempt to resolve the nothing at stake problem by punishing users who build on multiple forks. In 2014, Ethereum founder Vitalik Buterin created the \textit{Slasher} algorithm which deducts from users' deposits if they misbehave \cite{Buterin14}. However, the Slasher protocol requires users to be selected for participation in the lottery well ahead of time\footnote{See Ethereum's blog for more details \cite{Ethereum19}.}. This invites collusion because users can get together in advance of their turn and agree to launch a coordinated attack.

Other strategies simply punish any user who builds on the wrong fork. The protocol punishes all errors indiscriminately, making no effort to distinguish unlucky users from those who intentionally build on multiple chains. This will tend to discourage risk-averse users from extending the blockchain and magnify the power of more risk-prone adversaries. It could also enable adversaries to gang up on honest users. By manufacturing and communicating a set of ``bad" chains to their neighbors while refusing to pass on the correct chain, adversaries can cause honest peers to be punished any time they attempt to stake. This diminishes the wealth and influence of good actors while ensuring that dividends and power flow to the adversaries.

Monetary punishments cannot resolve the problems created by monetary rewards. They, too, create new issues in the process of solving the original.

\section{Incentives to Good Behavior}

The detrimental impact of cash incentives is not unique to blockchain. Research suggests that adding monetary incentives to a system does not always motivate desired behaviors. On the other hand, desired behaviors are often observed even when no cash motive is present.

\subsection{Day Care in Israel}

\noindent Monetary punishments are not sufficient to motivate desired behaviors. In a  well-known field study conducted during the 1990's, researchers imposed a fine on parents who were late to pick up their children from on a group of day care centers in Israel \cite{Gneezy00}. They found that parents arrived 30 minutes later on average when they were being charged. This effect persisted after the fine was lifted. These results contradicted the so-called ``deterrence hypothesis"---that introducing a penalty will reduce the penalized behavior and leave everything else unchanged.

But monetary incentives \textit{don't} leave everything else unchanged. According to the researchers, money ``change[s] the perception of the game." Putting a price on something doesn't just affect how much weight it carries in peoples' calculations. It restructures the calculations themselves, which can lead to unexpected and undesirable behaviors.

\subsection{Secondary School in Colombia}

\noindent Monetary rewards are also not sufficient to motivate desired behavior. In a recent World Bank experiment, researchers offered cash compensation to families in Colombia for sending their children to secondary school \cite{BarreraOsorio11}. They found that students in the reward program were 3\% more likely to attend school, but only 1\% more likely to re-enroll, indicating that the cash motive lost its power over time.

Moreover, non-participating students with siblings in the program were 3\% less likely to attend school and 7.3\% less likely to re-enroll. Rewarding one subset of the family effectively served as a punishment for the other members. This suggests that distributing rewards unevenly and infrequently may actually discourage unrewarded users from participating in a system.

These findings don't bode well for blockchain. Punishments might not deter malicious actors. The efficacy of rewards may diminish over time---and the positive impact on recipients may be dwarfed by the negative side-effects experienced by other members of the system.

\subsection{Blood Supply in the US and UK}

\noindent Cash incentives may not be necessary to motivate desired behavior. A study by Richard Titmuss, a former professor at the London School of Economics, compared the blood supply systems in the US and UK in the 1970's \cite{Titmuss70}. The British system relied on voluntary donors, whereas the American system was largely controlled by for-profit companies.

Titmuss discovered that the UK had a higher quantity and quality of blood transfusions than the US\@. His conclusions suggest that a nonmarket system based on donation may be more effective than a model which compensates users financially.

\subsection{Altruism in BitTorrent}

A similar effect has been observed in distributed computing applications. The BitTorrent service is a peer-to-peer file downloading system where users share bandwidth with each other to increase the overall speed of their own downloads \cite{Cohen03}. Each peer has a network of neighbors with whom it cooperates in a tit-for-tat manner---if a neighbor stops reciprocating, the other users will remove it from their network.

Recent research suggests that some peers give more bandwidth than they receive \cite{Piatek07}. Users with large bandwidth appear to share more than the minimal rate necessary for reciprocation. This acts like a ``progressive tax". The more users benefit from the system, the more additional resources they contribute to it.

Of course, the term ``altruism" is a bit misleading here. BitTorrent users, and even blood donors, are still motivated by self-interest. They are positioned to benefit from the systems they support. But these benefits are not easily converted into financial terms. The benefit of receiving a blood transfusion, for instance, can't be transferred or used to purchase other goods. Users are motivated by the intrinsic value of the system rather than by the wealth it enables them to accrue. This intrinsic motive produces behaviors that are beneficial to users and to the welfare of the system.

\section{Conclusion}
\label{sec:conclusion}

\noindent People need a reason to use and maintain a system, but that motive need not be explicit or cash-based. Monetary incentives often fail to motivate behavior in desired ways, and people willingly contribute to systems when no explicit incentive is provided. The utility of the system, not its auxiliary rewards, is often a powerful enough motive.

Incentive schemes as they currently stand are not an effective solution to blockchain's problems. They do not discourage minorities from gaining disproportionate power, as mining rewards encourage centralization in proof of work-based chains. They do not effectively motivate users to maintain the system, as validator rewards and punishments encourage honest users to act in a manner that invites attacks and undermines the stability of proof of stake-based chains. Further research is needed to develop blockchains that achieve these goals while avoiding the downsides of incentives.

\printbibliography

\end{document}